\documentclass[pre,twocolumn,showpacs,preprintnumbers,amsmath,amssymb]{revtex4}

\usepackage{amsmath}
\usepackage{amssymb}
\usepackage{graphicx}

\newcommand{\erf}{\mathop{\textrm{erf}}}
\newcommand{\var}{\mathop{\textrm{var}}}

\newcommand{\noref}[1]{}

\begin{document}
\preprint{submitted to \textit{Phys. Rev. E}}
\author{A. G. Rossberg}
\email{rossberg@uni-freiburg.de}
\homepage{http://www.fdm.uni-freiburg.de}
\author{K. Bartholom{\'e}}
\author{J. Timmer}
\affiliation{Zentrum f{\"u}r Datenanalyse und Modellbindung,
  Universit{\"a}t Freiburg, Eckerstr.\ 1, 79104 Freiburg, Germany}
\date{28 March 2002}

\title{Data driven optimal filtering for phase and frequency of noisy
  oscillations: application to vortex flowmetering}

\begin{abstract} 
  A new method for extracting the phase of oscillations from noisy
  time series is proposed.  To obtain the phase, the signal is
  filtered in such a way that the filter output has minimal relative
  variation in the amplitude (MIRVA) over all filters with
  complex-valued impulse response.  The argument of the filter output
  yields the phase.  Implementation of the algorithm and
  interpretation of the result are discussed.  We argue that the phase
  obtained by the proposed method has a low susceptibility to
  measurement noise and a low rate of artificial phase slips.  The
  method is applied for the detection and classification of mode
  locking in vortex flowmeters.  A novel measure for the strength of
  mode locking is proposed.
\end{abstract}

\maketitle

\section{Introduction}
\label{sec:introduction}

Several modern methods for time series analysis make explicit use of
the phase of measured oscillatory signals.  Examples are tests for
unidirectional \cite{pikovsky97:_phase_sync} or mutual
\cite{rosenblum96:_phase_synch_chaot_oscil} synchronization of chaotic
oscillators, based on accurate or noisy
\cite{tass98:_detec_lock_noisy} data, identification of the coupling
direction
\cite{rosenblum01:_detec_dir_coup,rosenblum02:_ident_coup_dir}, or
indicators \cite{parlitz96:_exper_phase_sync} for generalized
synchronization \cite{rulkov95:_gener_sync}.
Phase analysis has successfully been applied in neurology
\cite{schack00:bhattacharya01:}%
, cardiology
\cite{tass98:_detec_lock_noisy,%
stefanovska00:_rever_trans_synch_states_cardior_system,%
anishchenko00:_entrain_heart_rate_weak_nonin_forcin,%
rosenblum02:_ident_coup_dir%
}, ecology \cite{blasius00:_chaos_phase_synch_ecolog_system}, and
astronomy \cite{bhattacharya01:_applic_chrom} (for recent,
comprehensive reviews, see 
\cite{boccaletti02:_synchro_rev,pikovsky01:_synch}).

A number of methods have been proposed for the determination of the
phase.  Among these are (a) phase extraction from the argument of the
analytic signal \cite{gabor46:_theor}, from the convolution of the
signal with a Morlet wavelet
\cite{lachaux00:_study_singl_trial_phase_synch_activ_brain,%
deshazer01:_detec_phase_synch_chaot_laser_array%
}, or after complex demodulation or quadrature filtering
\cite{ktonas80:_instan_freq_phase}, (b) the angle of circulation of a
2D projection of the reconstructed phase-space trajectory
\cite{pikovsky01:_synch} or its time derivative
\cite{chen01:_weak_phase_coherent} around a point, and (c) linear
interpolation \cite{pikovsky97:_phase_sync} of phase between distinct
events marking the beginning new cycles \cite{pikovsky01:_synch}.

\newcounter{wish} \renewcommand{\thewish}{}
\newcommand{\wish}{\refstepcounter{wish}\thewish}

Although some rules for selecting the appropriate method for a given
system have been proposed \cite{pikovsky01:_synch}, the choice is not
always obvious.  The wish list for properties of the reconstructed
phase $\phi(t)$ includes: \wish\label{w:pi} a constant advance of
$2\pi$ per cycle, \wish\label{w:steady} a steady accumulation ($\dot
\phi(t)\approx \mathrm{const}.$), \wish\label{w:robust} accuracy in
the presence of measurement noise, \wish\label{w:unequivocal}
unambiguity with respect to $2\pi$ phase slips, and
\wish\label{w:function} a functional dependence on the current state
of the oscillator (locality).  Autonomy of the oscillator combined
with steady accumulation \noref{w:steady} and locality \noref{w:function}
of phase implies that $\phi(t)$ is the variable that corresponds to
the zero Lyapunov exponent of the system -- another desired property.

But only for perfectly periodic signals can all these wishes be
fulfilled.  For deterministic, chaotic oscillators the
linear-interpolation method (c) does often lead to a satisfactory
steady and local phase.  The problem of defining a steady, local phase
when the internal dynamics of a deterministic, chaotic oscillator
are known was treated rigorously in Ref.~\cite{josic01:_phase}.  But
as internal- and measurement noise become stronger, some temporal
averaging is required and locality in time\noref{w:function} has to
traded for accuracy\noref{w:robust} and/or
unambiguity\noref{w:unequivocal}.  In
Refs.~\cite{tass98:_detec_lock_noisy,%
schack00:bhattacharya01:,%
stefanovska00:_rever_trans_synch_states_cardior_system,%
anishchenko00:_entrain_heart_rate_weak_nonin_forcin,%
rosenblum02:_ident_coup_dir,%
blasius00:_chaos_phase_synch_ecolog_system,%
bhattacharya01:_applic_chrom%
}  the condition of
unambiguity \noref{w:unequivocal} was relaxed and only the cyclic phases
[$\phi(t)\mathop{\mathrm{mod}}2\pi$] of noisy oscillations were
used.  Thus data analysis was insensitive to phase slips, i.e., sudden
advances of the phase by $\pm 2\pi$, which may or may not be artifacts
of measurement noise.  Here we choose to be less demanding with
respect to locality\noref{w:function}, in favor of a
steady\noref{w:steady}, accurate \noref{w:robust} and, as much as
possible, unambiguous \noref{w:unequivocal} phase.

In order to identify a corresponding method of phase extraction,
notice that the computation of the analytic signal (or complex
demodulation) [method (a)] is generally recommended to be combined
with linear band-pass filtering of the desired oscillatory component
\cite{ktonas80:_instan_freq_phase,boashash92:_estim}.  The overall
effect is the application of a complex-valued, linear band-pass filter
\cite{deshazer01:_detec_phase_synch_chaot_laser_array}.  When the
method of delays is used for the phase-space reconstruction of the
angle-of-circulation method (b), the 2D projection is also equivalent
to complex-valued linear
filtering~
\cite{%
broomhead92:sauer93%
}, likewise for calculation of time
derivatives.

Finally, in the vicinity of a Hopf bifurcation, where
dynamics can be brought into Hopf normal form by a nonlinear
coordinate transformation (see, e.g., \cite{guh}), this transformation
is done in such a way that all contributions to dynamics which are
``non-resonant'' with the oscillation at the fundamental frequency are
eliminated.  In mere \emph{kinetic} terms this simply amounts to eliminating
higher harmonics and offsets, which can be achieved by complex, linear
band-pass filtering.  The $SU(2)$ symmetry of the Hopf normal-form
guarantees the steady accumulation of phase in the steady state, when
phase is measured as the angle of rotation around the origin.

Thus, a unified view on (a) and (b) is complex, linear band-pass
filtering.  It achieves steady accumulation\noref{w:steady} in a natural
way when the fundamental mode is isolated.  The results regarding
accuracy\noref{w:robust} and unambiguity\noref{w:unequivocal} depend on
the choice of the filter.  Since the concept of phase originates from
limit-cycle oscillations, which, in the transformed coordinates,
correspond to a motion on a circle, our idea for choosing the filter
here is to make the filtered signal move as close as possible to a
circle in the complex plane.  Roughly speaking, we consider the motion
on the circle as the signal and the deviations as noise and maximize
the signal to noise ratio (SNR) -- even though not all deviations are
actually due to measurement noise.  Since, with such a filter,
noise-induced excursions of the trajectory to the origin of the
complex plane are minimized, this is also a good way to reduce
ambiguities in the phase.  The maximization of the SNR is done not
only with respect to width and center frequency of the filter, but
with respect to the complete dynamics of its impulse response.  The
determination of the filter is non-parametric and data driven.


We proceed as follows.  Section~\ref{sec:theory} contains a
mathematical formulation of the ideas outlined above and lists some
implications.  In Section~\ref{sec:numerics} the method is applied to
simulated data, with special attention to the effect of filtering on
measured frequencies.  A practical application to vortex flowmeters is
discussed in Sec.~\ref{sec:applications}, where we also introduce a
novel measure for the strength of mode locking.

\section{Theory}
\label{sec:theory}

\subsection{MIRVA filters}
\label{sec:definition}

\subsubsection{Definition}

Let $x(t)$ be a real- or complex-valued stationary signal with
oscillatory components.  Denote by $z(t)$ the signal obtained from
$x(t)$ by linear filtering with a complex-valued filter with impulse
response $f(t)$, i.e., $z=f*x$.
Define $q$ as the nonnegative number such that
\begin{align}
  \label{def:q2}
  q^2=
  \frac{\var
    \left|
      z
    \right|^2}{
      \left<
      \smash{\left|
        z
      \right|^2}
    \right>^2
  }
  =
\frac{
    \left<
      \smash{\left|
        z
      \right|^4}
    \right>}{
    \left<
      \smash{\left|
        z
      \right|^2}
    \right>^2
  }
  - 1
  ,
\end{align}
where $ \left< \cdot \right> $ denotes the expectation value.  

Now, search for a filter $f$ such that the quantity $q$ given
by~(\ref{def:q2}) has a local minimum with respect to the filter. 
Such a filter $f$ minimizes the relative variance of the amplitude
(MIRVA) for the given signal $x(t)$.  The practical computation of
MIRVA filters is addressed in Appendix~\ref{sec:computational}.

For every MIRVA filter $f$, there is a two-parameter family of MIRVA
filters $f_{s,c}(t)=c f(t-s)$ with real $s$ and complex $c$.  Below we
shall always have a single member stand for the whole family without
saying.

\subsubsection{Example}
\label{sec:example}

\begin{figure}[t]
  \centering
  \includegraphics[width=\columnwidth,keepaspectratio]{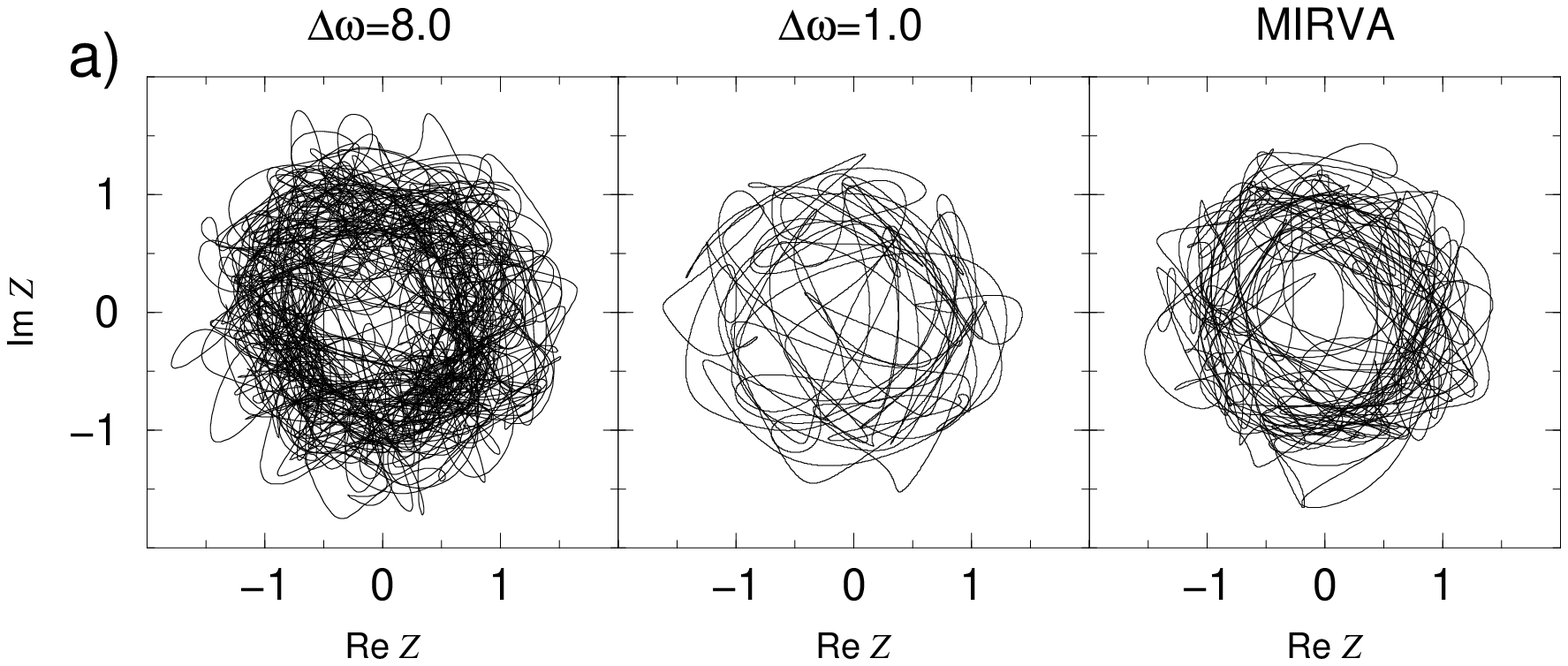}\\
  \includegraphics[width=\columnwidth,keepaspectratio]{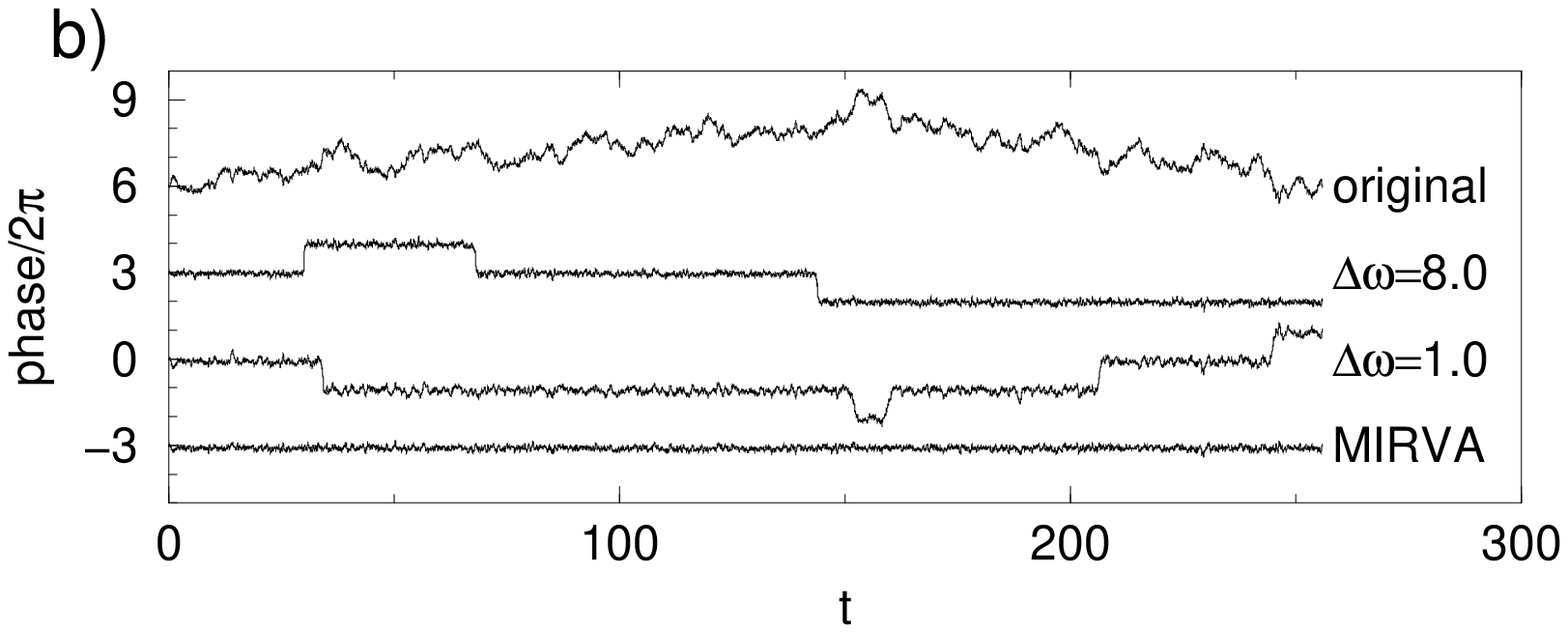}\\
  \caption{Three different filters applied to the same time series
    (\ref{phi-noise}).  a) The demodulated, filtered signals
    $Z(t)=z(t)\exp(-i\omega_0\,t)$.  b) The phase
    $\phi_0(t)-\omega_0\,t$ of the original signal and the relative
    phases $\phi(t)-\phi_0(t)$ obtained using the three filters.  See
    text for details.}
  \label{fig:example}
\end{figure}

As an example, assume that $x(t)$ is composed of a constant-amplitude
oscillation with phase fluctuations and white measurement noise,
\begin{align}
  \label{phi-noise}
  x(t)=\cos[\omega_0\,t+\phi_0(t)]+\eta(t),
\end{align}
where
\begin{align}
  \label{wiener-noise}
  \left<
    \smash{\dot \phi_0(t)\dot \phi_0(t')}
  \right>
  =2D\delta(t-t'),\quad
  \left<
     \eta(t) \eta(t')
  \right>
  =2G\delta(t-t').
\end{align}
This signal mimics narrow-banded limit-cycle and chaotic oscillations
in the vicinity of the fundamental frequency.  The larger the noise
strength $G$, the more difficult the determination of the phase
$\phi_0(t)$ from $x(t)$ becomes.  We simulate $x(t)$ with $D=1$,
$G=0.01$, and $\omega_0=20$ over an interval of length $T=256$.
Without any filtering, the SNR is zero.  Figure~\ref{fig:example}a
shows the demodulated signals $Z(t)=z(t)\exp(-i\omega_0\,t)$ for three
different complex filters $f(t)$.  The first two filters are of the
form
\begin{align}
  \label{gauss-filter}
  f(t)\sim\exp\left[i \omega_0\,t+ \frac{1}{2}(\Delta\omega\,t)^2\right].
\end{align}
One is a comparatively wide ($\Delta \omega=8.0$) band pass, the other
one is rather narrow ($\Delta \omega=1.0$).  The third is the MIRVA filter
obtained by minimizing (\ref{def:q2}).  It is approximated by
(\ref{gauss-filter}) with $\Delta \omega=2.9$.  As is shown in
Fig.~\ref{fig:example}b, both the narrow and the wide filter lead to
artificial phase slips.  Only when using the MIRVA filter is 
$\phi_0(t)$ faithfully tracked.

\subsubsection{Remarks on the minimization of  $q$}
\label{sec:basic}

Let $q_{\mathrm{min}}$ denote the value of $q$ attained at a local
minimum.  Since the operation of linear filtering defines a semi
group, $q_{\mathrm{min}}$ is also a local minimum of $q$ with respect
to further filtering of $z(t)$, i.e., for $q$ calculated with $z$
replaced by $z':=g*z$.  The minimum is attained when the filter $g$ is
the unit element of the semi group, the Dirac $\delta$-function.  As a
result one has
\begin{multline}
  \label{min-condition1}
  0=
  \left.\frac{1}{2}
      \frac{\delta q^2}{\delta g(\tau)}
  \right|_{g\equiv \delta}= \\
\frac{\left< \smash{\left| z(t) \right|^2} \overline{z}(t) z(t-\tau) \right>
  \left<
  \smash{
    \left|
      z(t)
    \right|^2}
  \right>-
  \left<
    \smash{\left|
      z(t)
    \right|^4}
  \right>
  \left<
    \smash{
      \overline{z}(t)
      z(t-\tau)
    }
  \right>}
{
  \left<
    \smash{\left|
      z(t)
    \right|^2}
  \right>^3
}  
\end{multline}
for all $\tau$.  In particular, when differentiating with respect to
$\tau$ at $\tau=0$ and taking the imaginary part,  it follows that 
\begin{align}
  \label{omegas}
  \frac{
  \left<
    \smash{\left|
      z
    \right|^4 \omega_i}
  \right>
  }{
  \left<
    \smash{\left|
      z
    \right|^4 }
  \right>
  }
  =
  \frac{
  \left<
    \smash{\left|
      z
    \right|^2 \omega_i}
  \right>
  }{
  \left<
    \smash{\left|
      z
    \right|^2 }
  \right>
  }
  = \omega_{\mathrm{mean}},
\end{align}
where $\omega_i$ is the instantaneous frequency defined by 
\begin{align}
  \label{def:omegai}
  \omega_i:=\mathop{\mathrm{Im}}
  \left\{
    \frac{\dot z}{z}
  \right\}
\end{align}
and $\omega_{\mathrm{mean}}$ is known as the \emph{mean frequency},
defined either by the last equation of (\ref{omegas}) or,
equivalently, as the ``center of mass'' of the power spectrum of the
signal $z$.  For the relation of the mean frequency to the \emph{phase
  frequency} (or average frequency) $\omega_{\mathrm{ph}}:=\left<
  \omega_i \right>$ see Sec.~\ref{sec:numerics}.

Often, signals contain oscillations at several different frequencies.
A systematic method to extract various oscillation frequencies has
been proposed in \cite{yalcinkaya97:_phase_of_chaos}.  When using
the concept of MIRVA filtering, distinct local minima of $q$ can be
identified with distinct oscillatory components of the signal.

\subsubsection{Special cases}
\label{sec:special_cases}

In two special cases the problem of finding MIRVA filters can be
discussed analytically: For perfectly periodic or quasi-periodic
signals there is, for every Fourier mode excited by the signal, a
MIRVA filters that extracts exactly this mode.  The filtered signal is
of the form $z(t)=\exp(i \omega t)$ and $q_{\text{min}}=0$.  This
holds true also if the signal is overlaid by any kind of noise.

For Gaussian, linear processes it is always possible to find filters
such that $\left< z^2 \right>=0$ and $\left< |z|^2 \right>\ne0$, e.g.,
by letting only Fourier modes with positive frequency pass.  Then $
\left< \smash{ \left| z \right|^4} \right>=2 \left< \smash{ \left| z
    \right|^2 }\right>^2$, $q=q_{\text{min}}=1$, and all these filters
are MIRVA filters.

The processes we are interested here are typically located between
these two poles: Noisy, nonlinear, periodic processes with some
fluctuations in the phase.  Thus we expect $0<q_{\text{min}}<1$.  When
solving the optimization problem $q^2=\mathrm{min}$ numerically with
time series of finite length $T$ (see Appendix~\ref{sec:computational}),
local minima with $q_{\text{min}}>1$ have also been found.  It is not
clear whether these persist in the limit $T\to\infty$.

\subsection{The phase  of MIRVA filtered signals}
\label{sec:phase}

\subsubsection{Definition and error estimates}

The main purpose of MIRVA filtering is to obtain the phase
\begin{align}
  \label{def:phi}
  \phi(t)=\int^t\omega_i(t')\,dt'\equiv \arg z
  \quad (\mathop{\mathrm{mod}} 2\pi)
\end{align}
of the oscillations extracted by the filter. 

Denote by $\eta(t)$ the contribution of measurement noise to $z(t)$,
the ``true'' value of $z(t)$ by $z_0(t):=z(t)-\eta(t)$, and the
``true'' phase by $\phi_0(t):=\int^t \mathop{\mathrm{Im}}\{\dot
z_0(t')/z_0(t')\}dt'$.  
Two kinds of errors in $\phi(t)$ caused by measurement noise can be
distinguished: deviations by multiples of $2\pi$, i.e.\ phase slips,
which are due to noise-induced excursions of $z(t)$ around the origin
and accumulate as time proceeds, and errors in the cyclic phase
$[\phi(t)-\phi_0(t)+\pi] \mathop{\mathrm{mod}} 2\pi$, which have a
finite correlation time.  The distinction is particularly sharp when
$q_\text{min}$ is small enough, so that the probability density for
values of $z(t)$ near zero is small, or, as we shall
consider now, when $q$ is small for general filters $f$.

An order of magnitude estimate for an upper bound to the rate of
noise-induced phase slips is given by
\begin{align}
  \label{slip_rate}
  p
  \left[
    \frac{|z|^2}{
      \left<
        \smash{|z|^2}
      \right>}
    =0
  \right]\cdot\Delta \omega,
\end{align}
where the first term denotes the probability density of ${|z|^2}/{
  \left< \smash{|z|^2} \right>}$ at zero and the second term denotes
the spectral width of the filter.  The first term typically decays
exponentially fast as $q^2$ decreases, while the relation between
$q^2$ and $\Delta \omega$ is only algebraic in general.  Hence,
minimizing $q^2$ is a good strategy to minimize phase slips. 

For the noise-induced error in the cyclic phase, an exact upper bound
can be obtained in the limit that $q^2$ is small.  For simplicity,
assume that the noise has undergone sufficient temporal averaging by
the filter $f$, so that the central limit theorem applies and $\eta(t)$ is
Gaussian.  Since $f$ is a complex band-pass filter,
$\left<\smash{\eta^2}\right>=0$. In order to derive an upper bound for
$\left<\smash{|\eta|^2}\right>$ from $q^2$, we assume the worst case, that is,
all variation in $|z|^2$ is due to $\eta$ only, while $|z_0|^2\equiv\mathrm{const.}=
1$ without loss of generality.  Since $\eta$ is independent of $z_0$,
Eq.~(\ref{def:q2}) can then be written as
\begin{align}
  \label{noise_in_q2}
  q^2=\frac{1+4\left<\smash{|\eta|^2}\right>+ 2
    \left<\smash{|\eta|^2}\right>^2
  }{1+2\left<\smash{|\eta|^2}\right>+\left<\smash{|\eta|^2}\right>^2}-1.
\end{align}
Solving for $\left<\smash{|\eta|^2}\right>$ yields
\begin{align}
  \label{eta}
  \left<\smash{|\eta|^2}\right>=
  (q^2-1)^{-1/2}-1=\frac{q^2}{2}+\mathcal{O}(q^4).
\end{align}
The corresponding noise-induced variance in $\phi$ is
\begin{equation}
  \label{phase-error}
  \begin{split}
    \var\phi=&
    \var\left[\arg(z_0+\eta)-\arg(z_0)\right]\\
    =&\var\left[ \arg \left( 1+\frac{\eta}{z_0}
      \right)\right]\\
    =&\var\left[\mathrm{Im}\left\{\frac{\eta}{z_0}\right\}\right]+\mathcal{O}(\left<\smash{|\eta|^2}\right>^2)\\
    =&\frac{\left<\smash{|\eta(t)|^2}\right>}{2}+\mathcal{O}(\left<\smash{|\eta|^2}\right>^2).
\end{split}
\end{equation}
In the general case, when $|z_0|$ is not constant, we get, from combining
(\ref{eta}) and (\ref{phase-error}),
\begin{align}
  \label{upper_var_phi}
  \var\phi\le \frac{q^2}{4} +\mathcal{O}(q^4).
\end{align}
Thus, minimizing $q^2$ is a good strategy to minimize noise-induced
errors in the measured cyclic phases.


\subsubsection{Phase diffusion}
\label{sec:diffusion}

Over long time intervals, $\phi(t)$ typically performs a random walk with
drift.  Thus, another important characteristic of the phase is its
diffusion coefficient
\begin{align}
  \label{def:D}
  D:=\lim_{T\to\infty}
  \frac{
    \left<
      \phi(t+T)-\phi(t)-\omega_\text{ph} T
    \right>
  }{
    2\, T
  }.
\end{align}
The estimation of $D$ from finite-length samples of $\phi(t)$ is
discussed in Appendix~\ref{sec:estimate}.

\subsection{Invariance with respect to filtering of $x(t)$}
\label{sec:invariance}

The MIRVA filtered signal $z(t)$ and the phase and frequency derived
thereof are invariant with respect to linear filtering of the original
signal $x(t)$ in the following sense: Let $y(t)$ be a signal obtained
from $x(t)$ by linear filtering, i.e. $y=h*x$, and let $f$ be a MIRVA
filter for $x$.  Then, a MIRVA filter for $y$ is, at least formally,
given by $f'=f*h^{-1}$, where $h^{-1}$ is the inverse of $h$ defined
by $h^{-1}*h=\delta$.  So the filtered signal $z=f*x=f'*y$ which satisfies
the minimization condition~(\ref{min-condition1}) is identical for $x$
and $y$.
For example, MIRVA filtering can be used to extract the phase and
frequency of an oscillatory signal which has been ``bleached''
\cite{theiler93}, i.e., filtered such as to make its power spectrum
white (see \cite{rossberg03:_frequency_robust_filtering}).

The concept of MIRVA filtering carries straightforwardly over to a
discrete-time representation of signals $x_i$, $z_i$ and filters
$f_k$, sampled at time intervals of length $\Delta t$.
In Ref.~\cite{rossberg03:_frequency_robust_filtering} the notion of a
\emph{topological frequency} of a time-discrete signal $x_i$ is
defined.  Roughly speaking, this is the rate of transitions of the
trajectory of the signal in a sufficiently high-dimensional delay
space through a particular kind of Poincar{\'e} section called a
\emph{counter}.  For the topological frequency, the invariance with
respect to linear filtering has been proven rigorously
\cite{rossberg03:_frequency_robust_filtering}.  When the modulus of the
signal $z_i$ obtained by MIRVA filtering a signal $x_i$ and its linear
interpolation have a nontrivial lower bound, i.e.,
$|l\,z_i+(1-l)\,z_{i+1}| > d > 0$ for $0\le l\le 1$, and when the
impulse response $f_k$ of the MIRVA filter decays sufficiently fast for
large $k$, then $\omega_{\text{ph}}$ obtained from $z_i$ and its
linear interpolation is (up to the sign) identical to a topological
frequency of the oscillations of the signal $x_i$.  A corresponding
counter can be obtained as follows: Assume that all significant
contributions to $f_k$ are within a range of $M$ consecutive delay
times.  Then the filter operation $f * x$ can be interpreted as a
projection from the $M$-dimensional delay space of $x_i$ into the
two-dimensional complex plane.  The counter is given by all points in
delay space which are projected onto the real, non-negative half axis.

\section{The effect of MIRVA filtering on measured frequencies}
\label{sec:numerics}


In order to illustrate the effects of MIRVA filtering on measured
frequencies, the method is applied to a numerical solution of the
noisy Stuart-Landau Equation (or Hopf normal form)
\begin{align}
  \label{stuart-landau}
  \dot A=(\epsilon+i \omega_l)A - (g_r+i\,g_i) |A|^2 A + \zeta,
\end{align}
where $A=A(t)$ represents the complex amplitude of an oscillator and
$\zeta(t)$ is complex-valued, Gaussian, white noise with correlations
\begin{align}
  \label{noise-strength}
  \left<
    \zeta(t) \zeta(t')
  \right>=0 
  \hbox{ and }
  \left<
    \zeta(t) \overline{\zeta}(t')
  \right>=4 G \delta(t-t').
\end{align}
In a certain sense, this system universally describes noisy
oscillations in the vicinity of a Hopf bifurcation
\cite{arnold98:_random_dynam_system}.  Assume the bifurcation to be
supercritical ($g_r>0$) and do a linear change of coordinates to set
$g_r=G=1$ and $\omega_l=0$ without loss of generality (even though, in
practice, $\omega_l\gg\epsilon$).  When $g_i\ne 0$, the phase
frequency,
\begin{align}
  \label{omega_ph}
  \omega_{\text{ph,A}}=-g_i\, \left(\epsilon+2 \mathcal{N}^{-1}\right)
\end{align}
(where $\mathcal{N}:=\pi^{1/2} \exp(\epsilon^2/4)
[1+\erf(\epsilon/2)]$, see, e.g.,
\cite{risken89:_fokker_planc_equat__chap12,rossberg03:_frequency_robust_filtering}),
calculated from $\omega_{i,A}=\mathrm{Im}\{\dot A/A\}$ directly
without filtering, differs from the corresponding mean frequency
\begin{align}
  \label{omega_mean}
  \omega_{\text{mean,A}}=-g_i \,
  \left[2
    \epsilon^{-1}+\epsilon-4\,
    \left(2 \epsilon+\epsilon^2 \mathcal{N}\right)^{-1}
  \right].
\end{align}
This is a direct consequence of the correlation between $\omega_{i,A}$
and $|A|^2$ ($ \left< \smash{\omega_{i,A} |A|^2} \right>/\left<
  \smash{|A|^2} \right>\ne \left< \smash{\omega_{i,A}}\right>$).

As an example, a simulation of $A(t)$ of length $T=10^6$ with
$\epsilon=2$, and $g_i=1$ is generated using Euler steps of length
$2^{-11}$.  For these parameters, $\omega_{\text{ph,A}}=-2.225$,
$\omega_{\text{mean,A}}=-2.899$, and the relative variance of the
unfiltered amplitude is $\left< \smash{|A|^4}\right>/ \left<
  \smash{|A|^2} \right>^2-1=(2 \mathcal{N}^2-2\epsilon\mathcal{N}-4) /
(\epsilon\mathcal{N}+2)^2=0.303$.

\begin{figure}[t]
  \centering
  \includegraphics[width=\columnwidth,keepaspectratio]{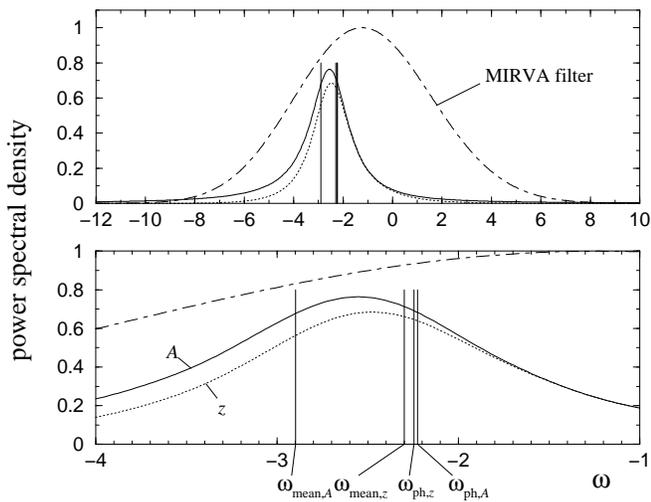}
  \caption{MIRVA filtering of the process
    $A(t)$ given by Eqs.~(\ref{stuart-landau},\ref{noise-strength})
    with $\omega_l=0$, $g_r=g_i=G=1$, and $\epsilon=2$.  The lower
    graph is a blow-up of the upper graph.  Both show the power
    spectral density of $A(t)$ (solid), of the filtered signal $z(t)$
    (dotted), and the characteristics of the MIRVA filter
    (dash-dotted).  The vertical lines indicate the locations of
    various frequencies associated with the process.}
  \label{fig:simulation}
\end{figure}

The MIRVA filter for the time series $x(t)=A(t)$ is calculated by the
indirect method described in Appendix~\ref{sec:computational}.
The filter reduces the relative variance of the amplitude to
$q_{\text{min}}^2=0.164$.  Figure~\ref{fig:simulation} shows the
characteristics of the MIRVA filter in comparison with the power
spectrum of the original signal $A(t)$ and the filtered signal $z(t)$.
The locations of phase- and mean frequency before filtering and after
filtering [$\omega_{\text{ph,z}}= -2.245(2)$, $\omega_{\text{mean,z}}=
-2.298(2)$] are also indicated.  The MIRVA filter is a rather wide
(half-width $2\,\Delta\omega\approx 6.5$), approximately symmetric
band-pass filter with a center frequency $\omega_c\approx -1.3$ below
the linear frequency $\omega_l=0$.  As a result, the phase frequency
of the filtered signal is also shifted to lower frequencies, but the
effect $\omega_{\text{ph,z}}-\omega_{\text{ph,A}}\approx -0.02
=\mathcal{O}(\Delta\omega^{-1}\mathcal{N}^{-1}\omega_c)$ (see
\cite{rossberg03:_frequency_robust_filtering}) is quite small.  

On the other hand, there is a pronounced shift in the mean frequency
by MIRVA filtering: the mean frequency approaches the phase frequency.
This is a generic effect of MIRVA filtering.  In the presence of a
negative correlation between amplitude and instantaneous frequency, as
found in our example, a filter that amplifies the signal when
$\omega_i$ is high and damps the signal when $\omega_i$ is low reduces
fluctuations in the amplitude and, at the same time, reduces the
correlation.  Since the phase frequency is only weakly affected by
MIRVA filtering, the mean frequency is shifted toward the phase
frequency.


\section{Application to vortex flowmetering}
\label{sec:applications}

\subsection{Background}
\label{sec:background}

\begin{figure}[b]
  \centering
  \includegraphics[width=0.4\columnwidth,keepaspectratio]{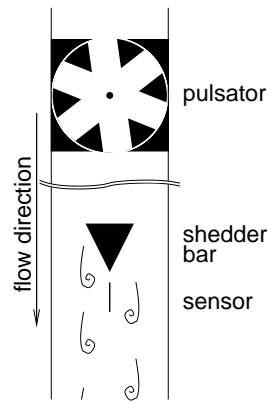}
  \caption{Schematic of the experimental setup  to record the sensor
    signal of a vortex flowmeter in pulsatile flow.  The shedder bar
    has a triangular cross section for efficient vortex generation.
  }
  \label{fig:setup}
\end{figure}

Next, an application of MIRVA filters to vortex flowmetering is
discussed.  Vortex flowmeters are widely used in the industry to
measure pipe flow.  The measurement principle makes use of the
phenomenon of the von-K{\'a}rm{\'a}n vortex street.  Behind a shedder
bar inserted normal to the flow in a pipe, a regular chain of vortices
is formed, rotating alternatingly left and right.  The volume flow
through the tube can be determined from the frequency of vortex
formation.  In the device used here, a piezoelectric sensor sensitive
to transversal flow, which is inserted downstream behind the shedder
bar, is used to detect the vortices passing by.  A common problem of
vortex flowmetering is mode locking of the vortex oscillations to
pulsations in the flow.  The second-order statistics (power spectra)
of the sensor signal and the bias on the flow measurement in the
presence of mode locking have been thoroughly investigated
\cite{%
mottram91:amadi-echendy93:peters98:%
}.
But it seem possible that, by processing the sensor signal with a
stronger focus on the nonlinear dynamics of the system, a better
control of mode-locking can be achieved.

Here we describe the analysis of time series recorded in an experiment
designed to simulate the problem of detecting mode locking in an
industrial context, using only the sensor signal.  Methods that have
been proposed to detect mode locking from univariate time series are
the analysis of the map of subsequent period lengths of the
oscillation (angle-of-return-times-map)
\cite{%
janson01:janson02:janson02:%
} and the application of the established bivariate methods on
pairs of time series extracted from the univariate series by
filtering \cite{stefanovska00:_spatial_sync}.  We go along the lines
of the second approach, making it more powerful by applying MIRVA
filters to separate the signals.

The setup of the experiment is sketched in Fig.~\ref{fig:setup}.
Pulsations of the pipe flow were generated by a rotating cylinder with
three bores orthogonal to the cylinder axis, which is inserted into
the pipe in such a way that, by the rotation, the flow is periodically
blocked.  This pulsator is driven by an electric motor.  The sensor
signal of a commercial flowmeter, which was mounted about 40 pipe
diameters downstream from the pulsator, was recorded.  Estimates of
the pulsation rate $\nu_{\text{puls}}$ and the frequency of vortex
formation $\nu_{\text{vort}}$ were available on-site, while recording
the time series.  Reynolds numbers were $\mathcal{O}(10^5)$ and the
flow was highly turbulent.  As a result, both inherent and measurement
noise contribute substantially to the signal.  Details of the
experimental setup will be reported elsewhere.
Below, two experimental time series labeled as A and B are discussed.
Both were recorded from the sensor of the vortex flowmeter at
$2\,\mathrm{kHz}$ over $250\,\mathrm{s}$.

\begin{figure}[t]
  \centering
  \includegraphics[width=\columnwidth,keepaspectratio]{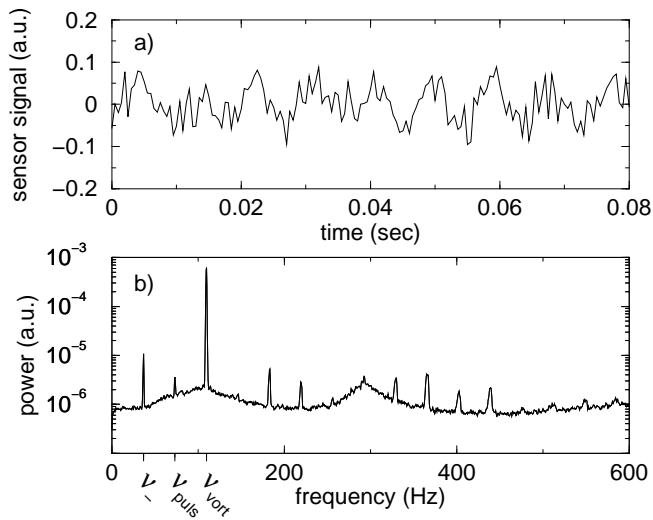}
  \caption{A representative section (top) and the power spectral
    density (bottom) of Signal A, which was recorded in the experiment
    sketched in Fig.~\ref{fig:setup}.}
  \label{fig:signal}
\end{figure}

\begin{figure}[t]
  \centering
  \includegraphics[width=\columnwidth,keepaspectratio]{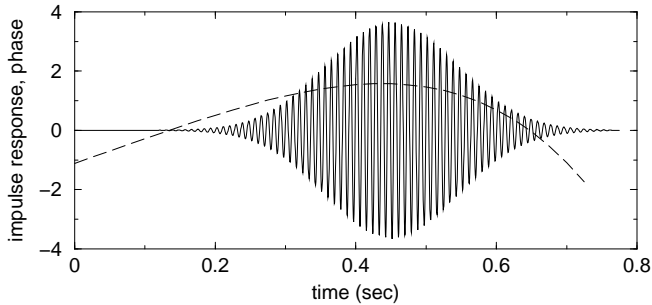}
  \caption{The impulse response $f(t)$ of the MIRVA filter calculated
    for Series A at the frequency $\nu_\text{vort}$.  Solid:
    $\mathrm{Re}\{f(t)\}$; dashed: phase of $f(t)$, relative to an
    oscillation at constant frequency $\omega_0/2
    \pi=109.64\,\mathrm{Hz}$, i.e., $\arg \left[f(t)\exp(-i\omega_0
      t)\right]$.  The overall offset in time is an accidental choice
    of the search algorithm.}
  \label{fig:response}
\end{figure}

\subsection{Series A: hard lock-in}
\label{sec:seriesA}

\subsubsection{Extraction of phases}

When recording series A, the flow rate was adjusted such as to obtain
$\nu_{\text{vort}}\approx 110\,\mathrm{Hz}$ and the pulsation frequency
was set to $\nu_{\text{puls}}\approx \frac{2}{3} \nu_{\text{vort}}$.  The
time series was analysed to determine the strength of the expected
$2:3$ lock-in.  In Fig.~\ref{fig:signal} a representative section of
Series A and the power spectrum are shown.  The oscillations at
$\nu_{\text{vort}}$ can clearly be seen.  Since the pulsations
themselves do not produce any transversal flow, there is only a weak
signal at $\nu_{\text{puls}}$, presumably due to slight asymmetries in
the setup.  By the nonlinear interaction of vortex street and
pulsation, flow oscillations at $\nu_-=\nu_{\text{vort}}-
\nu_{\text{puls}}$ are excited.  These contain significant transversal
components and can clearly be seen in the power spectrum.  The power
spectrum also reveals several other oscillatory components in the
signal.  Some of these are nonlinearly generated, others are of
unknown origin.

\begin{figure}[t]
  \centering
  \includegraphics[width=\columnwidth,keepaspectratio]{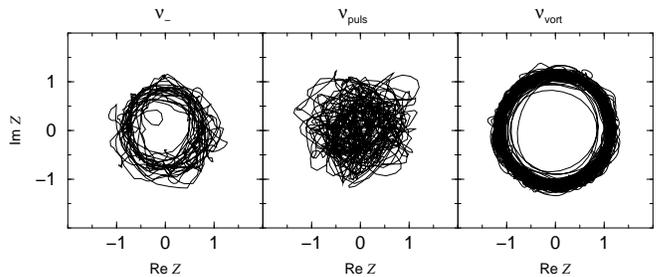}
  \caption{The trajectories of  demodulated, MIRVA filtered signals
    $Z_i$ (see Appendix~\ref{sec:regularization}), obtained from
    Series~A at the indicated frequencies.  The corresponding values
    of $q_\text{min}$ are $0.41$ ($\nu_-$), $0.81$
    ($\nu_\text{puls}$), and $0.13$ ($\nu_\text{vort}$).}
  \label{fig:demods}
\end{figure}

\begin{figure}[b]
  \centering
  \includegraphics[width=\columnwidth,keepaspectratio]{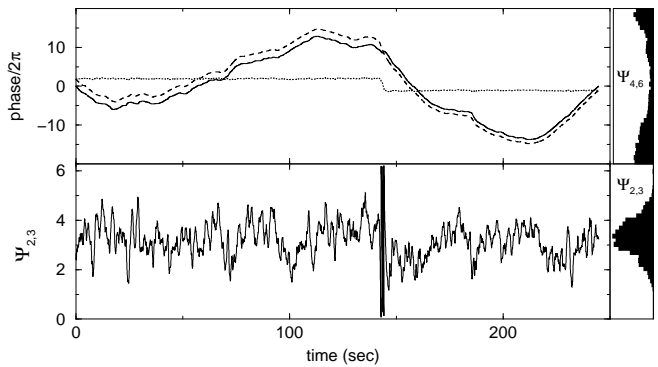}
  \caption{Phases, extracted from Series A. Large upper panel: the
    unwrapped phases $3\phi_-(t)-\omega_0 t$ (solid) and
    $\phi_\text{vort}(t)-\omega_0 t$ (dashed), where
    $\omega_0/2\pi=109.654\,\mathrm{Hz}$ (nominal value);
    $\phi_{2,3}=3\phi_{-}-\phi_\text{vort}$ (dotted).  Large
    lower panel: the cyclic relative phase $\Psi_{2,3}$.  Small
    panels: Empirical distributions of $\Psi_{4,6}$ and $\Psi_{2,3}$}
  \label{fig:phases}
\end{figure}

The impulse response of the MIRVA filter for the $110\,\mathrm{Hz}$
vortex oscillations ($\nu_\text{vort}$) is shown in
Fig.~\ref{fig:response}.  It was calculated by the indirect method
described in Appendix~\ref{sec:computational} using a down-sampling
factor $h=50$, $M=30$ sampling points, a demodulation frequency
$\omega_0/2\,\pi=109.643\,\mathrm{Hz}$, and a regularization by
$m=10$th order polynomials.  The overall Gaussian shape of the impulse
response of the MIRVA filter can clearly be seen.  But the filter has
additional structure.  The oscillation frequency of the response
function decreases with time (see the phase in
Fig.~\ref{fig:response}).  The reason for this particular phase
dynamics is not clear by now.  As can be seen from the trajectory of
the demodulated filtered signal $Z_i$ shown in Fig.~\ref{fig:demods}
(right), the phase of the vortex oscillations is always well defined.

From the construction of the pulsator it is clear that the flow
pulsations have a well defined phase: Each passage of a bore along the
pulsator inlet (or outlet) defines one puls.  But the signal-to-noise
ratio of the oscillations at $\nu_{\text{puls}}$ is too low to derive
unequivocal phase information.  As is shown in Fig.~\ref{fig:demods}
(center), the MIRVA filtered signal at $\nu_{\text{puls}}$ repeatedly
reaches the origin of the complex plane.

In contrast, the phase of the oscillations at $\nu_-$ is much better
defined (see Fig.~\ref{fig:demods}, left).  Since the signal-to-noise
ratio is smaller at $\nu_-$ than at $\nu_\text{vort}$, the MIRVA
filter at $\nu_-$ is about 
8 times more narrow in Fourier space than the MIRVA filter at
$\nu_\text{vort}$.  Use of the MIRVA filter (or some approximation) is
critical for phase extraction at $\nu_-$.  Here, straightforward boxcar
filtering of a region in Fourier space containing the $\nu_-$ peak
(see, e.g., Ref.~\cite{stefanovska00:_spatial_sync}) would be
insufficient.

The phase $\phi_-$ of the oscillations at $\nu_-$ can be used to
determine the phase $\phi_\text{puls}$ of the pulsator.  
By the physical interpretation of the
oscillation at $\nu_-$ as a nonlinear excitation, one has the relation
\begin{align}
  \label{phim}
  \phi_-=\phi_{\text{vort}}-\phi_{\text{puls}},
\end{align}
that yields $\phi_{\text{puls}}$ for known $\phi_-$ and $\phi_{\text{vort}}$.

\subsubsection{Relative phases and symmetry}

From the $2:3$  mode locking, one expects that the
relative phase $\phi_{2,3}$, given by 
\begin{align}
  \label{def:phinm}
  \phi_{n,m}:=n\, \phi_\text{vort} - m\, \phi_\text{puls},
\end{align}
changes only little over time.  For hard mode locking, it fluctuates around
a constant value.  With both hard and soft lock-in, the cyclic
relative phase $\Psi_{2,3}$ defined by
\begin{align}
  \label{def:psi}
  \Psi_{n,m}:=\phi_{n,m} \mathop{\mathrm{mod}} 2 \pi
\end{align}
has an uneven distribution.  Generally, one expects the distribution
to be increasingly sharper localized to a single value as mode locking
becomes stronger.  The synchronization index defined in
Ref.~\cite{rosenblum01:_phase_data_analysis} as
\begin{align}
  \label{def:gamma}
  \gamma_{n,m}^2:=
  \left<
    \cos \Psi_{n,m}
  \right>^2
  +
  \left<
    \sin \Psi_{n,m}
  \right>^2,
\end{align}
with expectation values estimated by temporal averaging, was therefore
proposed as a measure for the strength of mode locking or, more
general, phase locking.  Absence of mode locking is indicated by
$\gamma_{n,m}=0$, hard coupling by $\gamma_{n,m}=1$.  In our
experiment, we encounter the particular situation that vortices and
pulsation have opposite symmetries with respect to transversal
reflection.  Thus, dynamics is equivariant already with respect to a
shift of $\phi_\text{vort}$ by $\pi$, rather than $2\pi$.  Ideally, one
would therefore always expect $\gamma_{2,3}^2=0$, with or without mode
locking.  In order to take this degeneracy into account,
$\gamma_{4,6}^2$ should be used instead of $\gamma_{2,3}^2$.

\subsubsection{Interpretation of extracted phases}

The evolution of the measured values for $\phi_-$ and
$\phi_\text{vort}$, and of the relative phase
$\phi_{2,3}=2\,\phi_\text{vort}-3\,\phi_\text{puls}=
3\,\phi_- -\phi_\text{vort}$ is shown in
Fig.~\ref{fig:phases}.  Since the definition of MIRVA filters leaves
the overall delay of the filtered signal undetermined, the relative
delay of $\phi_-$ and $\phi_\text{vort}$ has to be adjusted in a
reasonable way.  We choose the delay such that $\gamma_{2,3}$ becomes
maximal ($\gamma_{2,3}$ turns out not to vanish, see below).  From the
evolution of $\phi_{2,3}$ it appears that the vortex oscillations
contain only a single phase slip at about 150 seconds into the time
series.  But upon closer inspection, it appears more plausible to
account the phase slip to an error in measuring $\phi_-$: As
expected for this case, the difference in $\phi_{2,3}$ before and
after the slip is to a good accuracy $6\pi$.  For slips in
$\phi_\text{vort}$, any other multiple of $\pi$ would have been
possible as well.  Furthermore, the slip occurs just at the moment
when the demodulated MIRVA signal at $\nu_-$ (Fig.~\ref{fig:demods},
left) goes through the small loop reaching toward the coordinate center.  It
appears that the MIRVA filter is too narrow for the comparatively low
pulsation frequency at this moment.  In fact, the phase slip
disappears when wider filters are used -- at the price of obtaining
new artificial phase slips at other times.  In conclusion, the data
indicate that there is not a single real phase slip. Lock-in is hard over
the full $250\,\mathrm{s}$ sampling time.

The relevant synchronization index $\gamma_{4,6}^2=0.16$ is much
smaller than one would expect for hard mode locking (see also the
distributions of $\Psi_{4,6}$ in Fig.~\ref{fig:phases}).  Use of a
synchronization index based on Shannon entropy
\cite{tass98:_detec_lock_noisy} yields a similar result.  Even when
the transversal reflection symmetry was strongly broken, the then
relevant synchronization index $\gamma_{2,3}^2=0.62$ would be rather
low.  But, as can be seen from Fig.~\ref{fig:histogram} below, the
symmetry is only weakly broken.  A natural explanation for the
discrepancy between the synchronization index and the phase-slip
statistics is to assume that most of the fluctuations in $\Psi_{2,3}$,
respectively $\Psi_{4,6}$ (Fig.~\ref{fig:phases}; $\Psi_{4,6}=2
\Psi_{2,3}\mathop{\mathrm{mod}}2\pi$), are due to measurement noise,
and not intrinsic to the vortex dynamics.  This view is compatible
with the upper bound derived for the variance due to noise in
Sec.~\ref{sec:phase}.  From $q_-:=q_{\text{max}}=0.41$ at $\nu_-$ and
$q_\text{vort}:=q_{\text{max}}=0.13$ at $\nu_\text{vort}$, one gets the
approximate upper bound $3^2 q_{-}^2/4+q_{\text{vort}}^2/4=0.38$ for
the variance contributed to $\Psi_{2,3}$ by measurement noise, while
the total variance is $\var \Psi_{2,3}=0.48$.  It appears that, in
some situations, a characterization of mode locking by phase-slip
statistics is less susceptible to measurement noise than
characterizations based only upon cyclic phases, i.e., quantities such
as $\phi_{\text{puls/vort}} \mathop{\mathrm{mod}} 2\pi$ or -- computable
thereof -- $\Psi_{4,6}$.

\subsection{Series B: soft lock-in}
\label{sec:seriesB}

\begin{figure}[tb]
  \centering
  \includegraphics[width=\columnwidth,keepaspectratio]{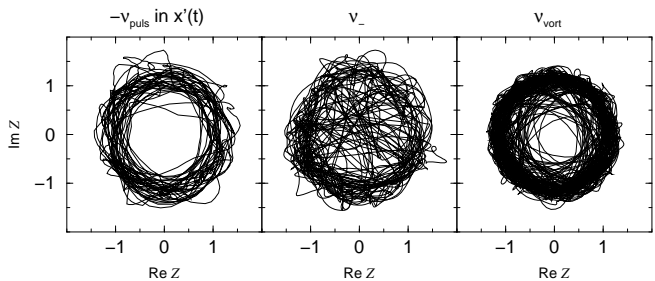}
  \caption{The trajectories of  demodulated, MIRVA filtered signals
    $Z_i$ (see Sec.~\ref{sec:regularization}), obtained from Series~B
    and the quotient signal $x'(t)$, defined in Sec.~\ref{sec:seriesB},
    at the indicated frequencies.  The corresponding values of
    $q_\text{min}$ are $0.30$ ($-\nu_\text{puls}$ in $x'(t)$), $0.39$
    ($\nu_-$), and $0.22$ ($\nu_\text{vort}$).}
  \label{fig:demodsB}
\end{figure}

\begin{figure}[b]
  \centering
  \includegraphics[width=\columnwidth,keepaspectratio]{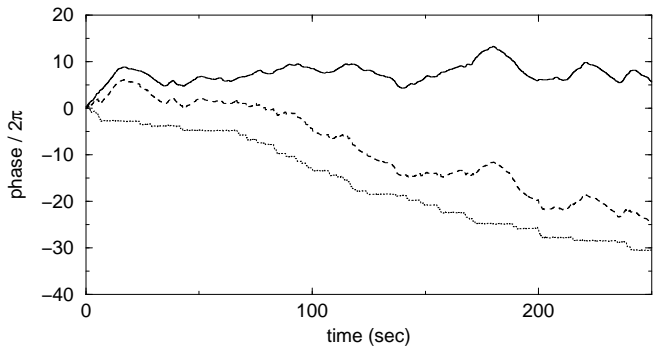}
  \caption{Unwrapped phases, extracted from Series B:
    $2\phi_\text{puls}(t)-\omega_0 t$ (solid) and
    $\phi_\text{vort}(t)-\omega_0 t$ (dashed), where
    $\omega_0/2\pi=110.697\,\mathrm{Hz}$ (nominal value); the relative
    phase $\phi_{1,2}=\phi_\text{vort}-2\phi_\text{puls}$
    (dotted).}
  \label{fig:phasesB}
\end{figure}

\begin{figure}[tb]
  \centering
  \includegraphics[width=\columnwidth,keepaspectratio]{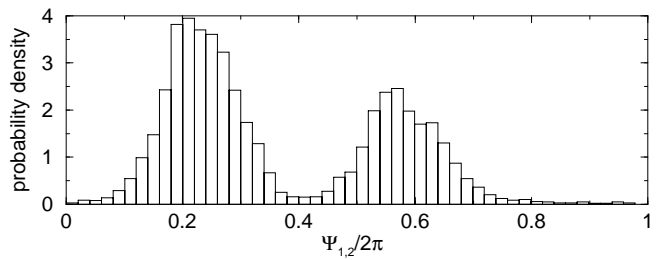}
  \caption{Empirical distribution function of the cyclic relative
  phase $\Psi_{1,2}$ obtained from Series B.  The double peak is a
  result of the transversal reflection symmetry of the experimental
  system (Fig.~\ref{fig:setup}).}
  \label{fig:histogram}
\end{figure}

\subsubsection{Extraction of phases}

Series B was recorded with $\nu_\text{vort}\approx110\,\mathrm{Hz}$
and $\nu_\text{puls}\approx55\,\mathrm{Hz}$ and $1:2$ mode locking is
expected.  Similar as for Series~A, MIRVA filtering readily extracts
the unwrapped phase $\phi_\text{vort}$ of the oscillation $\sim\exp(i
\phi_\text{vort})$ at $\nu_\text{vort}$ (Fig.~\ref{fig:demodsB},
right).  We assume that, as for Series A, the oscillations $\sim
\exp(i \phi_\text{puls})$ due to pulsation alone are much weaker than
the oscillations $\sim\exp(i \phi_\text{vort}-i \phi_\text{puls})$
excited by nonlinear interaction of vortices and pulsation.  Thus, the
nonlinear excitation dominates the oscillations at
$\nu_\text{puls}\approx \nu_-:=\nu_\text{vort}-\nu_\text{puls}$.  In
contrast to Series A, MIRVA filtering at $\nu_-$ does not lead to an
unequivocal phase (Fig.~\ref{fig:demodsB}, center).  It appears that
this is due to phase slips in $\phi_\text{vort}$, which broaden the
range of relevant frequencies at $\nu_-$ and, as a result, worsen the
signal-to-noise ratio.

By making use of the MIRVA filtered signal $z_\text{vort}(t)$ of the
vortex oscillations, $\phi_\text{puls}$ can nevertheless be
extracted from the signal.  In variation of a method proposed in
Ref.~\cite{gasquet97:_variab_freq_dem}, a new, complex-valued time
series $x'(t):=x(t)/z_\text{vort}(t)$ is constructed from the original
signal $x(t)$.  The overall effect of this transformation is to shift
all oscillations by $\nu_\text{vort}$ to negative frequencies.  The
oscillations that were at $\nu_-$ are now at $\nu_- -
\nu_\text{vort}=-\nu_\text{puls}$. They are of the form $\sim\exp(-i
\phi_\text{puls})$, i.e., they do not depend on the phase of the
vortices.  Now, MIRVA filtering $x'(t)$ at $-\nu_\text{puls}$ yields the
desired unequivocal phase information (Fig.~\ref{fig:demodsB}, left).
As is shown in Fig.~\ref{fig:phasesB}, the $\phi_\text{vort}$
follows $\phi_\text{puls}$, but several phase slips
occur. 

The histogram of the cyclic relative phase $\Psi_{1,2}$ reveals the
two preferred phase angles which are due to transversal reflection
symmetry.  Since the symmetry is weakly broken, their separation is
not exactly $\pi$.  Again, it is not clear to what degree the
broadening of the distribution of $\Psi_{1,2}$ is due to measurement
noise and to what degree to internal noise.  

\subsubsection{Quantification of the degree of mode locking}

In order to quantify the degree of mode locking independent of
measurement noise, a characterization that depends only on the
long-term dynamics of the phases would be useful.  Such a measure
is, for example,  given by $\rho_{1,2}$, with
\begin{multline}
  \label{rho}
  \begin{aligned}
    \rho_{n,m}:=& \frac{ n^2 D \left[ \phi_\text{vort} \right] +
      m^2 D \left[ \phi_\text{puls} \right] - D \left[
        \phi_{n,m} \right] }{ 2\, m^2 D \left[ \phi_\text{puls}
      \right]
    }\\
    =&\lim_{T\to\infty}
  \end{aligned}
  \\
  \frac{n\, \mathop{\mathrm{cov}} \left[
      \phi_\text{vort}(t+T)-\phi_\text{vort}(t) ,
      \phi_\text{puls}(t+T)-\phi_\text{puls}(t) \right] }{2\,m\,
    \mathop{\mathrm{var}} \left[
      \phi_\text{puls}(t+T)-\phi_\text{puls}(t) \right] },
\end{multline}
where $D[\cdot]$ stands for the diffusion coefficient of the specified
phase variable.  $\rho_{n,m}$ measures in how far the response
oscillator (here vortices) follows phase fluctuations of the drive
oscillator (pulsation).  When $\phi_\text{vort}$ and
$\phi_\text{puls}$ evolve independently, $\rho_{n,m}=0$ for
all $n$, $m$.  In the case of hard $n:m$ lock-in, as was found for
Series A, $D\left[ \phi_{n,m} \right]=0$, $n^2 D \left[
  \phi_\text{vort} \right] = m^2 D \left[ \phi_\text{puls}
\right]$, and $\rho_{n,m}=1$.  Weak mode coupling interpolates between
these two extremes.  Values of $\rho_{n,m}$ outside the range $[0,1]$
are possible in principle but unphysical in the situation of direct,
unidirectional coupling.  Since only the long-term dynamics of the phases is
taken into account, rather long time series are required to obtain
reproducible values of $\rho_{n,m}$.  For Series B we find, using the
estimator given by Eq.~(\ref{def:Dhat}) with $\tau=12.5\, \mathrm{s}$,
$D\left[ \phi_{1,2} \right]=0.4(1)\,\mathrm{s}^{-1}$, $D\left[
  \phi_\text{vort} \right]=2.4(6)\,\mathrm{s}^{-1}$, and
$4\,D\left[ \phi_\text{puls}\right]=2.5(6)\,\mathrm{s}^{-1}$,
resulting in $\rho_{1,2}\approx0.9$.  The empirical value of
$\rho_{1,2}$ is stable over a wide range in $\tau$.  Of course,
$\rho_{n,m}$ could not be used when the frequency of the drive
oscillator was perfectly stable, i.e., when $D \left[
  \phi_\text{puls} \right]=0$.  A detailed analysis of the measure
$\rho_{n,m}$ and its interpretation is yet to be worked out.

\section{Conclusion}
\label{sec:conclusion}

MIRVA filtering was introduced as a new method for extracting the
phase of oscillations from noisy time series.  It was argued that the
phase so obtained is, among other favorable properties, particularly
robust to noise and linear filtering of the signal.  Detailed
directions for computing MIRVA filters numerically are given in
Appendix~\ref{sec:computational}.  In a numerical case study it was
demonstrated that MIRVA filtering introduces only little bias to the
phase (or average) frequency.  A new synchronization index has been
proposed, which is designed to be robust to noise if MIRVA filtering
is used.

By applying MIRVA filtering to the signal of a vortex flowmeter, we
showed that the method can be used to obtain well defined phases from
oscillatory time series even under unfavorable conditions such as
strong internal and measurement noise.  The phases were used to
investigate the strength of mode locking.  As another application, the
MIRVA filtered signal was used for a data driven demodulation
technique in Sec.~\ref{sec:seriesB}.  Limitations to phase extraction,
which remain even when MIRVA filters are used, have been
addressed.


\acknowledgments

The authors express their gratitude to F.~Buhl and ABB Automation
Products~GmbH for providing the vortex-flowmeter data, to P.~Riegler and
Y.-K.~The for insightful discussions, and to the German
Bundesministerium f{\"u}r Bildung und Forschung (BMBF) for generous
supported (grant 13N7955).

\appendix

\section{Computation of MIRVA filters}
\label{sec:computational}

\subsection{Main algorithms}
\label{sec:algorithms}

In practical applications time series $x_i$ ($i=1,...,N$),
sampled from $x(t)$ at evenly spaced discrete times $t=i\Delta t$ are
given.  The MIRVA filters have to be estimated from this data.  Here,
two methods are proposed.  The first method is more appropriate for
short time series, the second is more efficient when $N$ is large.
With both methods, the impulse response $f_j$ of the filters is
restricted to a finite length $M$ ($j=1,\ldots,M$).

\subsubsection{Direct method}

When using the first method, the convolution
\begin{align}
  \label{convolution}
  z_k=(x*f)_{k+M}=\sum_{j=1}^{M} f_j x_{k+M-j}
\end{align}
(for $k=1,\ldots,N-M+1$) is calculated directly, and the expectation
values in Eq.~(\ref{def:q2}) are estimated as averages over $k$.  An
iterative search algorithm is used to find the MIRVA filter $f_j$ with
$q^2=q^2_{\text{min}}$.

\subsubsection{Indirect method}

The second method makes use of the fact that $q$ depends on $x(t)$
only through its second and fourth moment.  It is often more efficient
than the first method but, as a trade off, entails systematic errors of
the order $\mathcal{O}(M/N)$ in the estimation of the moments of
$z_i$.  For notational convenience, we define the second and fourth
moments $x_i$ in the time-discrete representation as
\begin{align}
  \label{fourth}
  c_{i\,j\,k\,l}&=
  \left<
    x_{\tau-i}\,
    x_{\tau-j}\,
    \overline{x}_{\tau-k}\,
    \overline{x}_{\tau-l}
  \right>\\
  \intertext{and}
  \label{second}
  c_{i\,j}&=
  \left<
    x_{\tau-i}\,
    \overline{x}_{\tau-j}
  \right>
\end{align}
with arbitrary $\tau$.
These expectation values are estimated by averaging over time and making
use of symmetries, e.g., $c_{i\,j}=\overline{c}_{j\,i}=c_{i+k\,j+k}$.

Equation~(\ref{def:q2}) now reads
\begin{align}
  \label{q2fast}
  q^2=
\frac{
\sum_{i\,j\,k\,l} c_{i\,j\,k\,l} f_i f_j \overline{f}_k \overline{f}_l
}{
  \left(
    \sum_{i\,j} c_{i\,j} f_i \overline{f}_j
  \right)^2
}-1
\end{align}
with all sums running over $1,\ldots,M$.  Thus, while the computation
of the moments takes time of order $\mathcal{O}(N\,M^3)$, the time
required for the optimization itself is independent of $N$.  Besides,
derivatives of $q^2$ with respect to $f_i$ are calculated at little
additional cost, and can be provided to the optimization algorithm
to help finding a MIRVA filter $f_i$.

\subsection{Down-sampling, demodulation, and regularization}
\label{sec:regularization}

In order not to introduce artificial restrictions of the search space
for $f_i$, the duration of the impulse response, i.e. $M \times \Delta
t$ should be of the order of the phase coherence time of the
oscillation.  This time can easily cover several hundred oscillation
periods.  Computation time depends critically on $M$. To keep $M$ low
and make the computation feasible, it is therefore advisable not to
work with the raw time series $x_i$ but to perform a demodulation and
a down-sampling step prior to the main calculation.  For the
calculations discussed in
Secs.~\ref{sec:numerics},\ref{sec:applications}, instead of $x_i$, the
demodulated time series $X_j$ given by
\begin{align}
  \label{def:X}
  X_j:=\sum_l K_l\, x_{h j+l}\, \exp[-i\,(h j+l)\,\omega_0\,\Delta t],
\end{align}
were used with a symmetric, triangular smoothing kernel $K_l$ at a
width of two time the down-sampling factor $h$.  The demodulation
frequency $\omega_0$ should be set to a value close to frequency of
the targeted oscillations.

To see the effect of this transformation, notice that for stationary,
discrete-time processes the value of $q$ defined by Eq.~(\ref{def:q2})
with $z$ replaced by $Z=F*X$ is for any filter $F_k$ identical to the
value obtained with $z=f*x$, provided
\begin{align}
  \label{re-f}
  f_k=\sum_l K_l\, F_{(k+l)/h}\, \exp(i\, k\,\omega_0\, \Delta t)
\end{align}
and $F_{k}:=0$ for non-integer $k$ by convention: It is not
difficult to verify that $Z_{j}=z_{hj}\exp(-ihj \omega_0 \Delta t)$,
independent of $K_l$.  

As a result, every MIRVA filter $F_k$ for $X_j$ leads by
Eq.~(\ref{re-f}) to the approximate MIRVA filter $f_k$ for $x_i$.  The
approximation is good if the interpolation (\ref{re-f}) of $F_k$
defined by $K_l$ is reasonable.

The MIRVA filters $F_k$ found for typical experimental data are more
or less deformed variants of Gaussian filters
$F_k\approx\exp(-\frac{1}{2}(k-M/2)^2 h^2 \Delta \omega^2 \Delta t^2)$
with bandwidth $\Delta\omega$.  The linear interpolation for $f_k$
given by the triangular $K_l$ is good if $h\, \Delta \omega \,\Delta
t\ll1$.  In practice, this requires filter lengths of at least
$M \approx 15-30$.

Experimental time series are often not long enough to yield faithful
estimates for all $\mathcal{O}(M^3)$ independent elements of
$c_{i\,j\,k\,l}$.  This problem can be overcome by a regularization of
$F_k$.  In our calculations, we restricted the filters $F_k$ to the
family $F_k=\exp(P_m(k))$, with $m$-th order polynomials $P_m$.

\subsection{A guide to choosing appropriate parameters}
\label{sec:guide}

The following procedures were used to find appropriate values for the
demodulation frequency $\omega_0$ and the down-sampling factor $h$,
which determines the duration of the filter's impulse response $h M \Delta t$.
The sampling rate $\Delta t$ is assumed to be given and $M$ is
restricted to a small range by computational limitations.

After an initial guess, the value of $\omega_0$ was set to the value
of the empirically found phase frequency $\omega_{\text{ph}}$ of $z_i$
in an iterative process.  In order to adjust $h$, the envelope $|F_k|$
of the computed MIRVA filter was investigated.  When $h$ is too large,
$|F_k|$ has a sharp peak and vanishes for all other values.  When $h$
is too small, most weight of the filter is concentrated near the end
points $F_1$ and $F_M$.  By inspection one finds $F_1\approx \pm
i\,F_M$, i.e., the MIRVA filter with a constraint in the
filter length approximates a simple 2D delay embedding. $h$ is
adjusted accordingly.

A systematic procedure for finding good values for the polynomial
order $m$ has not been developed yet.  But with $m\approx 6-10$
results generally depend little on the precise value.

\subsection{Convergence and side minima}
\label{sec:convergence}

In Sec.~\ref{sec:theory} it was proposed to identify local minima of
$q$ with distinct oscillatory components of the signal.  The
structure of the search space is therefore of immediate
theoretical interest.  In fact, with  long enough time series sampled
from a typical signal, the numerical search algorithms used here
consistently and effortlessly reach local minima located in a small
set of well separated points in the space of all filters, irrespective
of the -- randomly chosen -- starting points.  A unique minimum is
typically singled out when using demodulation and down sampling, since
this effectively implies a pre-selection of the frequency range of
interest.  With shorter time series, however, these minima split into
large clusters of several side minima.  In order to cope with
these artificial multiplicities, only those local minima were accepted
for selecting MIRVA filters which where found three times within a
series of minimization runs with random starting points, without
previously finding any point with a lower value of $q$.

\section{Remarks on the estimation of $D$}
\label{sec:estimate}

We discuss a method to determine the diffusion coefficient $D$ defined
by Eq.~(\ref{def:D}) from samples $\phi(t)$ of finite length ($0\le t
\le T$).  First $\omega_\text{ph}$ is estimated as
$\hat\omega_\text{ph}=[\phi(T)-\phi(0)]/T$.  Then $D$ can be estimated
by
\begin{align}
  \label{def:Dhat}
  \hat D_\tau :=
  \frac{
    \int_0^{T-\tau} 
    \left[
      \phi(t+\tau)-\phi(t)-\hat \omega_\text{ph} \tau
    \right]^2 dt
  }{
    2\,\tau\,(T-\tau)\,(1-\tau/T)
  },
\end{align}
where $0<\tau < T$.  The last factor in the denominator compensates
for the loss of statistical degrees of freedom by the estimation of
$\omega_\text{ph}$ as $\hat \omega_\text{ph}$.  When assuming
$\phi(t)$ to perform a random walk with constant drift, it is
straightforwardly verified that $\hat \omega_\text{ph}$ is a maximum
likelihood estimator and $\left<\smash{ \hat D_\tau}\right>=D$.  Under
the same assumption, the variance of $\hat \omega_\text{ph}$ is $2D/T$
and
\begin{multline}
  \label{D_Cov}
  \mathop{\mathrm{cov}}( \hat D_{k\, T}, \hat D_{l\, T})=\\
  D^2\cdot\frac{l[ 6{\left( 1 - k \right) }^2k - 
        2\left( 1 - k \right) \left( 1 + 3k^2 \right) l - 
        \left( 1 - 4k \right) l^2 ]}{3\,
      k\, {\left( 1 - k \right) }^2{\left( 1 - l \right) }^2}
\end{multline}
for $0<k\le l\le 1/2$ (the last expression was obtained with the help
of symbolic computer algebra).  In particular,
\begin{align}
  \label{D_variance}
  \var \hat D_{l\, T}=D^2\cdot
    \frac{l\,\left( 4 - 11\,l + 4\,l^2 + 6\,l^3 \right) }
    {3\,{\left( 1-l \right) }^4},
\end{align}
which increases monotonically with $0<l\le1/2$.  For good estimates of
$D$, the parameter $l$ should be chosen as small as possible but large
enough to justify the assumption of a random walk over times $l\,T$.
The estimator $\hat D_\tau$ can be slightly improved by using linear
combinations with different $\tau$.  For example, the variance of
\begin{align}
  \label{def:Ds}
  \hat D^\prime_{\tau}:=\frac{3}{2}\hat D_\tau-\frac{1}{2}\hat D_{2
    \tau}
\end{align}
is about $10\%$ smaller than of $\hat D_\tau$, as is verified using
Eq.~(\ref{D_Cov}).


\end{document}